\documentclass[twocolumn,amssymb,nofootinbib]{revtex4-2}

\usepackage{latexsym,amsfonts,amsmath,amssymb,graphics}
\usepackage{dsfont}
\usepackage[dvips]{graphicx}
\usepackage{mathrsfs}  %\mathscr{..} 
\usepackage{bbm}       %\mathbbm{..} 
\usepackage{slashed}   %\slashed{a} 
\DeclareMathAlphabet{\mathpzc}{OT1}{pzc}{m}{it}

\newcommand{\eq}[1]{\begin{equation}#1\end{equation}}
\newcommand{\eqs}[1]{\setlength\arraycolsep{2pt}
\begin{eqnarray}#1\end{eqnarray}}
\newcommand{\refer}[1]{(\ref{#1})}

\def\nn{\nonumber}
\def\bea{\begin{eqnarray}}
\def\eea{\end{eqnarray}}
\def\beq{\begin{equation}}
\def\eeq{\end{equation}}

\newcommand{\anti}[1]{\overline{#1}}
\newcommand{\derp}[2]{\frac{\partial #1}{\partial #2}}   %Cz?stkowa
\newcommand{\deru}[2]{\frac{{\rm d} #1}{{\rm d} #2}}     %Zwyk?a
\newcommand{\derf}[2]{\frac{\delta #1}{\delta #2}}       %Funkcjonalna
\newcommand{\cI}[0]{{\mathcal{I}}}
\newcommand{\cJ}[0]{{\mathcal{J}}}

\newcommand{\cO}[0]{{\mathcal{O}}}

\newcommand{\sD}[0]{{\mathscr{D}}}

\newcommand{\ep}{\epsilon}

\newcommand{\myAuthor}
{Adrian Lewandowski}

\newcommand{\myEmail}
{lewandowski.a.j@gmail.com}

\newcommand{\myAbstract}
{
Three-loop counterterms for the Standard Model (SM) revealed that the matrix of anomalous dimensions ($\gamma$) of quarks is divergent in the $d \to 4$ limit unless a carefully chosen non-Hermitian square-root of $Z$ matrix is used in the textbook formula for $\gamma$.  Here, an alternative prescription is given, which expresses $\gamma$ and $\beta$ functions directly in terms of counterterms (instead of $\sqrt{Z}$ and conventional `bare couplings') and produces finite results. In the SM, this prescription \emph{automatically} reproduces results obtained previously by adjusting $\sqrt{Z}$. 
}

\newcommand{\myRevtexTitle}
{ 	
  Divergences in anomalous dimension matrices of quarks at three loops:\\
       Explanation and simple solution
}

\newcommand{\myRevtexAffiliation}
{ 
Department of Theoretical Physics, Faculty of Fundamental Problems of Technology,
\\ Wrocław University of Science and Technology, 50-370, Wrocław, Poland
}

\newcommand{\myGenerateRevtexHeader}
{
  \title       { \myRevtexTitle }
  \author      { \myAuthor }
  \email       { \myEmail }
  \affiliation {\myRevtexAffiliation}
  \begin{abstract} \myAbstract \end{abstract}
  \pacs{12.60.Fr,1480.Ec,14.80Va}
  \maketitle
}

\begin{document}

%// For REVTEX version
\myGenerateRevtexHeader

%// For non-REVTEX version
%\myGenerateNonRevtexHeader

%OLD (LONGER) ABSTRACT:
%Three-loop calculations of renormalization group (RG) coefficients performed in recent years, revealed that anomalous dimensions of quarks, defined in terms of Hermitian square-roots of $Z$ matrices,  are divergent in the $d \to 4$ limit. The standard solution to this problem boils down to multiplication of $\sqrt{Z}$ matrices by carefully chosen unitary factors that also diverge in this limit, and give finite anomalous dimensions. In the present paper, it is shown that these divergences are caused by  taking square-roots of $Z$ matrices in the first place. An alternative prescription is given, which expresses anomalous dimensions in terms of $Z$ matrices (instead of $\sqrt{Z}$). Similarly, beta functions are expressed in terms of so obtained anomalous dimensions and complete counterterms for appropriate vertices (rather than conventional `bare couplings'). The resulting RG coefficients are proved to be finite. On more practical side, the prescription was tested against the three-loop counterterms for the Standard Model available in the literature and it automatically reproduces beta functions and anomalous dimensions obtained previously with the aid of the aforementioned unitary factors. Thus, the proposed prescription provides a direct mapping from the standard counterterms to finite RG functions.   

%\section{Introduction} 
\noindent {\bf Introduction.}
The dependence of renormalized correlation functions on renormalization scale $\mu$, which may appear to a newbie particle physicist as an additional parameter in quantum version of a given classical model, provides a powerful method of predicting  terms in perturbation expansion that have not been calculated yet 
\cite{Kastening}. 
Indeed, apart from an extra parameter, the quantum theory possesses also an additional (purely quantum) symmetry known as the (perturbative) renormalization group equation (RGE). Unlike most of other global symmetries, RGE mixes different orders of perturbation calculus, and thus allows for extraction of (some of) the higher order terms. Moreover, RGE gives an important consistency check for multi-loop calculations, as many of the terms in perturbation expansion are fixed by lower order RGE coefficients.             

While RGE is by now a standard textbook material, it may seem surprising at first that the  standard formula for the anomalous dimension, i.e. 
\eq{\label{Eq:gammaNaive}
{\gamma_F} =
\left(\sqrt{Z_F}\right)^{-1}
\mu\deru{}{\mu}
\sqrt{Z_F}\,,
}
leads at three-loop order to  $\gamma_F$ matrices in the Standard Model (SM), which still contain poles at $d=4$ in the $\overline{{\rm MS}}$ scheme of Dimensional Regularization, if  
$\sqrt{Z_F}$ is the Hermitian square-root of a matrix 
that renormalizes kinetic terms (see Eq. \refer{Eq:LagrWeylOg} below) of Weyl fields that create/annihilate quarks  \cite{3LSM}. The same problem has been observed in two Higgs doublet models (2HDMs) \cite{3L2HDM}. 
Moreover, as factoring out $\sqrt{Z_F}$ is a part of the standard definition of `bare couplings', problems with anomalous dimensions affect also the beta functions for Yukawa matrices of quarks. 
It was shown in the original papers \cite{3LSM,3L2HDM} that a unitary matrix $\mathcal{U}$ can be found (which by itself also diverges in the $d\to 4$ limit) such that the replacement 
$\sqrt{Z_F}\mapsto \mathcal{U}\sqrt{Z_F}$ in Eq. \refer{Eq:gammaNaive}, as well as in the definition of `bare Yukawa matrices', leads to finite anomalous dimensions 
and finite beta functions for Yukawa matrices. 
It was also emphasized in Ref. \cite{3L2HDM} that, even without the unitary factors, the poles at $d=4$ only affect beta functions of unphysical parameters, such as the Yukawa matrices, while beta functions of flavor-invariant quantities  in the quark sector are finite. More recently, Ref. \cite{Ambi} proved that these divergent anomalous dimensions and beta functions still lead to a finite RG flow of renormalized correlation functions. At the same time a generic  prescription for  finite (`flavor improved') beta and gamma functions is given in \cite{Ambi}, however these functions are defined in terms of additional diagrams involving external fields.         

Still, the RGE coefficients in a generic renormalizable model are most easily expressible in terms of these unphysical Yukawa matrices, and the like. 
Moreover, the RGE is an intrinsic feature of renormalized correlation functions, and as such has, in principle, nothing to do with counterterms. 
Indeed, even in renormalization schemes without divergent counterterms, the RGE 
controls terms in perturbation expansion, see e.g. \cite{PiguetSorella}. 
Thus, the necessity to search for additional divergent unitary factors, or to calculate extra diagrams of a spurion field, to get finite RGE coefficients, seems unnatural (not to mention working with divergent RGE coefficients).  
The purpose of the present paper is therefore to provide a solution to the problem of divergences in RGE coefficients, which is more natural 
and easier to use than approaches proposed so far in the the literature.

The RGE coefficients are usually calculated from counterterms, rather than from  renormalized correlation functions, especially in Dimensional Regularization (DimReg). This is, however, not the cause of the problems discussed in this paper, as RG-invariance simply reflects regulator-independence of renormalized correlation functions which, because of purely dimensional reasons, still remember that the logarithmic divergences existed. 
But, unlike plain counterterms,  
square-roots of $Z_F$ factors are completely unnatural creatures from the point of view of perturbative calculations. While such square-roots allow for an introduction of `bare fields' and `bare couplings', working with the bare objects is usually inconvenient and unnecessarily complicates problems one is trying to solve. For instance, the proof of renormalizability of non-Abelian gauge theories in gauge-symmetry-preserving regularizations is simpler if one abandons bare fields altogether, as then the nonrenormalization theorems for gauge fixing terms in linear gauges have more natural form \cite{ZJorig,ZJbook}. In fact, to the best of my knowledge, there is no regularization that is at the same time  \emph{consistent} and  \emph{automatically preserving} chiral gauge symmetries, see e.g. \cite{Bon} and references therein. Therefore, renormalization is always an additive procedure, even in `multiplicatively renormalizable' theories, and there is really no obvious definition of `bare fields' and `bare couplings'. 
One should keep this fact in mind, even though 
 practical multi-loop  calculations (in particular, the ones discussed here) are usually done in DimReg with the chiral-gauge-symmetry-preserving naive prescription for $\gamma^5$, what requires verification that its inconsistencies does not (yet) affect the result. 
%However, problems with Hermitian square-roots of $Z_F$ discovered in  \cite{3LSM,3L2HDM} indicate that `bare fields' (and couplings) are less natural (or at least more difficult to find) than one initially thought. 

Despite their unnaturalness, the $\sqrt{Z_F}$ factors play a crucial role in a textbook derivation of RGE. Consider the Lagrangian density that generates renormalized one-particle-irreducible (1PI) correlation functions (i.e. includes all the relevant counterterms, summed up into $(Z_F)_{ab}$, $(Z_S)_{jk}$, $\mathscr{Y}_{ j ab}$ etc.) 
\eqs{\label{Eq:LagrWeylOg}
\mathcal{L}_{tree}+\mathcal{L}_{c.t.}&=&
i (Z_F)_{ab}\anti{\chi}^a\anti{\sigma}^\mu\partial_\mu\chi^b
+\frac{1}{2} (Z_S)_{jk} \eta^{\mu\nu}\partial_\mu\phi^j\partial_\nu\phi^k
\nn\\
{}&{}&
-
\frac{1}{2}\phi^j\left(
\mathscr{Y}_{j ab}\chi^{a}\chi^{b}+
\mathscr{Y}^{*}_{j ab}\anti{\chi}^{a}\anti{\chi}^{b} 
\right)
+\ldots
}
Here $\chi^b$, $b=1,2,\ldots$ are Weyl fermions with different flavors/colors (spinor indices are supressed), $\phi^j$ are Hermitian scalar fields, 
$(Z_F)_{ab} = \delta_{ab}+\cO(\hbar)$ is a Hermitian matrix, while 
$(Z_S)_{ij} = \delta_{ij}+\cO(\hbar)$ is a real symmetric matrix. The ellipsis represents all other operators in the Lagrangian, in particular the quartic scalar couplings as well as interactions involving the gauge fields. The coefficients 
$\mathscr{Y}_{j ab}$ are symmetric in  $a\leftrightarrow b$ indices, and in DimReg in $d=4-2\epsilon$ dimensions they read 
\eq{ \label{scrY}
 \mathscr{Y}_{j ab}=\mu^{\epsilon}\left( {Y}_{j ab} + \mathfrak{Y}_{j ab}\right)\,, 
}
where ${Y}_{j ab}$ are renormalized Yukawa matrices, and 
$\mathfrak{Y}_{j ab}=\cO(\hbar)$ 
is the total counterterm for the $\chi\chi\phi$ vertex. 
Once the square-roots of $Z_{F,S}$ matrices are factored out from 
$\mathscr{Y}_{j ab}$ to define the conventional `bare couplings' 
${Y}^{B}_{j ab}$, one sees that the renormalized 1PI correlation functions 
depend on renormalized couplings and $\mu$ only through the bare couplings and 
the square-roots of $Z_{F,S}$ matrices. Moreover, factors $Z^{-1}_{F,S}$ from the propagators, cancel the square-roots of $Z_{F,S}$ from vertices connected by these propagators; thus what remains is a single $\sqrt{Z_{F,S}}$ factor for each external line of a 1PI function. This observation is enough to write the RGE for that function 
(see e.g. \cite{ZJbook} for the derivation of the Callan–Symanzik equation along these lines).   

As said, however, neither $\sqrt{Z_{F,S}}$ nor ${Y}^{B}_{j ab}$ are natural 
objects calculated in perturbation theory. By contrast, $Z_{F,S}$ and $\mathscr{Y}_{j ab}$ naturally follow from Feynman diagrams. In fact, one of the ancillary files \cite{3L2HDMfiles} for Ref. \cite{3L2HDM} provides $Z_{F,S}$ and $\mathscr{Y}_{j ab}$ (and their counterparts for some other vertices) at 3-loop order for the SM and 2HDMs as a ready-to-use {Mathematica} package. It is therefore desirable to have a prescription for RGE coefficients, as well as a derivation of the RGE itself, that operates on natural entities $Z_{F,S}$, $\mathscr{Y}_{j ab}$, etc. 

The remainder of the paper is organized as follows. First, a natural prescription for the RGE coefficients $\beta$ and $\gamma$ is given, and finiteness of the resulting $\beta$'s and $\gamma$'s is shown in general, as well as in the special case of 3-loop counterterms \cite{3L2HDM,3L2HDMfiles} for the SM. Next, a natural  derivation of the RGE, from which this prescription originates, is provided. 

\vspace*{3pt}
 
%\section{Prescription}
\noindent {\bf Prescription.}
Consider once again the Lagrangian density \refer{Eq:LagrWeylOg} which generates renormalized 1PI correlation functions. Let 
$\{g^C\}$ be the set of all independent renormalized parameters of a model. In particular, among $g^C$ there are real and imaginary parts of independent entries of renormalized Yukawa matrices $Y_{jab}$, as well as gauge couplings, quartic couplings and mass(-square) parameters (in non-Landau gauges, the gauge fixinig parameters must also be included in this set).     
I want to stay as close as possible to perturbative calculations, and therefore prefer not to talk about running couplings. Instead, I track only the explicit dependence of correlation functions (and counterterms) on $\mu$ and the renormalized parameters $g^C$. In this approach, running couplings are nothing more than solutions of the resulting RGE for renormalized 1PI correlation functions via the method of characteristics. 
    
Introducing a (yet to be found) beta function ${\beta}^C$  for each independent parameter $g^C$, 
one can define the differential operator \footnote{
This operator could (but not necessarily should) be identified with $\mu\deru{}{\mu}$, i.e. the total derivative with respect to $\mu$ with $g^C$ treated as a running coupling. I did such an identification in Eq. \refer{Eq:gammaNaive} to conform to the standard notation used in the literature.}
\eq{\label{scrD}
\mathscr{D}=\mu\derp{}{\mu} 
+
{\beta}^C\derp{}{g^C}\,.
}  
Beta functions and anomalous dimensions can now be easily obtained by solving (in perturbation theory) the 
following system of linear equations (flavor indices are suppressed and matrix multiplication is  used instead, $\mathscr{Y}_{k}$ is a matrix with matrix elements $\mathscr{Y}_{k ab}$)
\eqs{
\sD Z_F &{=}&   Z_F \gamma_F  + \gamma_F^{\dagger}  Z_F  \label{DZF}\,, \\
\gamma_F^{\dagger} &{=}& \gamma_F \,, \label{Herm}\\
\sD Z_S &{=}& Z_S \gamma_S  + \gamma_S^{\rm T}  Z_S\,,  \label{DZS} \\
\gamma_S^{\rm T} &{=}& \gamma_S \,, \label{Symm}\\
\sD \mathscr{Y}_{j} &=& {} 
\mathscr{Y}_{j} \gamma_F + \gamma_F^{\rm T} \mathscr{Y}_{j}
  +(\gamma_S)^k_{\ j} \mathscr{Y}_{k},   \label{DZY}
}
together with analogous equations for other vertices and gauge fields. 
I have decided to separate out the Hermiticity (symmetry) conditions for 
anomalous dimensions of Weyl fermions (respectively, Hermitian scalars) as they serve a different purpose than remaining equations. 
%Eqs.  \refer{DZF}, \refer{DZS} and \refer{DZY}. 
It will be shown below that any beta functions and anomalous dimensions that obey 
Eqs.  \refer{DZF}, \refer{DZS} and \refer{DZY} (together with their counterparts for remaining fields and vertices) guarantee that the renormalized 1PI generating functional satisfies the RGE. By contrast, Hermiticity (symmetry) conditions ensure that the resulting RGE coefficients have finite limits when the regulator is removed. Before discussing finiteness, one first has to realize that, since both sides of Eqs. \refer{DZF} (resp. \refer{DZS}) are explicitly Hermitian (resp. symmetric) and 
$Z_{S,F}=\mathds{1}+\cO(\hbar)$, the Hermitian (symmetric) part of $\gamma_F$ ($\gamma_S$) is uniquely fixed by these equations at every finite order of perturbative expansion. To get the $n$-loop contribution to this Hermitian (symmetric) part one needs only the $n$-loop contribution to $Z_{S,F}$ and 
$(n-1)$-loop contributions to beta functions. 
Similarly, Eq. \refer{DZY} determines in perturbation 
theory the $n$-loop contribution to the beta function 
for a Yukawa matrix $(Y_j)_{ab}=Y_{jab}$ (cf. Eq. \refer{scrY})
\eq{
  \beta_{Y_j} \equiv \beta^C \derp{}{g^C} {Y_j} \,,
} 
provided that $n$-loop contributions to $\mathscr{Y}_j$ and $n$-loop contributions to $\gamma_{S,F}$  are known. Thus, Eqs. \refer{DZF}-\refer{DZY}, supplemented by their counterparts for other field and vertices, have a unique solution in perturbation theory. It should be also stressed that 
\refer{DZF}-\refer{DZY} are valid not only in DimReg. In particular, in DimReg 
$Z_{S,F}$ are strictly dimensionless and therefore the partial $\mu$-derivative in \refer{scrD} does not contribute to the left hand side of Eqs. \refer{DZF} and \refer{DZS}. By contrast, in mass-independent schemes based on  (some sort of) cutoff regularization, $Z_{S,F}$ do depend explicitly on $\ln(\Lambda/\mu)$. In both classes of regularizations, 
$\gamma_{S,F} = \cO(\hbar)$. Just like $Z_{S,F}$, the counterterm $\mathfrak{Y}_{j ab}$ in Eq.\refer{scrY} has no $\mu$-dependence in DimReg, but unlike $\gamma_{S,F}$ beta functions in DimReg (because of the factor $\mu^{\epsilon}$) have nonzero tree-level contributions that vanish only in the $d\to4$ limit,    

To prove finiteness of the resulting $\beta$ and $\gamma$ coefficients, it is now enough to realize that Eqs. \refer{DZF}, \refer{DZS} and \refer{DZY} 
are structurally identical with the RGEs for the (formfactors of) corresponding 
\emph{renormalized} 1PI correlation functions of fermions and scalars. These  formfactors, just like  $Z_{S,F}$ and $\mathscr{Y}_{j}$,  have the form 
$\mathds{1}+\cO(\hbar)$ and, respectively, ${Y}_{j}+\cO(\hbar)$. Therefore 
one can uniquely express beta and gamma functions in terms of renormalized (i.e. finite) correlation functions provided the Hermiticity/symmetry conditions 
\refer{Herm}/\refer{Symm} are imposed. 

I have explicitly verified the correctness of the above prescription using the three-loop SM counterterms from   Refs. \cite{3L2HDM,3L2HDMfiles}. Since these counterterms are available as Mathematica 
files, solving Eqs. \refer{DZF}-\refer{DZY} is a simple exercise in Mathematica 
programming. \footnote{A full version of calculations reported here can be found in a supplementary Mathematica notebook \cite{MNB}.} 
In Ref. \cite{3L2HDM} the tree-level Lagrangian reads (I discuss here only the SM case)  
\eq{\nn
 \mathcal{L}^{d=4}_{tree} = - \left( \overline{Q_L} \tilde{\Phi} Y^{(u)} u_R 
                          +    
                          \overline{Q_L} \Phi Y^{(d)} d_R \right) +h.c.+\ldots,                          
}
with the quark $SU(2)$-doublet ${Q_L}$, quark $SU(2)$-singlets $u_R$ and $d_R$, as well as  the Higgs doublet $\Phi$ and 
$\tilde{\Phi}\equiv i \tau_2 {\Phi}^{*}$ 
with a Pauli matrix $\tau_2$. Expressing the components of Dirac fields 
via the corresponding Weyl fields ($u_R\simeq\bar{\chi}_u$, 
$d_R\simeq\bar{\chi}_d$, $\overline{Q_L}\simeq\bar{\chi}_Q$) one gets
\eq{\nn
 \mathcal{L}^{d=4}_{tree} = - \left(\bar{\chi}_Q\tilde{\Phi} Y^{(u)} \bar{\chi}_u 
                          +    
                           \bar{\chi}_Q \Phi Y^{(d)} \bar{\chi}_d \right) +h.c.+\ldots,                          
}
where flavor, color and spinor indices are suppressed. In particular, 
$Y^{(u)}$ and $Y^{(d)}$ correspond to certain submatrices of $Y_j^*$, cf. Eqs. \refer{scrY} and \refer{Eq:LagrWeylOg}.
Ref. \cite{3L2HDMfiles}, in addition to $Z_{u_R}=Z_{\chi_u}^*$, $Z_{d_R}=Z_{\chi_d}^*$,  
$Z_{Q_L}=Z_{\chi_Q}$ and $Z_\Phi$,  contains also matrices  
$Z_{Qu\Phi}$ and $Z_{Qd\Phi}$ that 
are submatrices of ${Y}_{j ab}^* + \mathfrak{Y}_{j ab}^*$, cf. Eq. \refer{scrY}, i.e.
\eq{\nn
 \mathcal{L}_{tree}+ \mathcal{L}_{c.t.}
 \supset - \mu^{\ep}\left(\bar{\chi}_Q\tilde{\Phi} Z_{Qu\Phi} \bar{\chi}_u 
                          +    
                           \bar{\chi}_Q \Phi Z_{Qd\Phi} \bar{\chi}_d \right) +h.c.                         
}
Because of (unbroken) $SU(2)$ gauge symmetry, $Z_\Phi$ is effectively a single (real) parameter, and so is $\gamma_\Phi$. Eqs. \refer{DZF}-\refer{DZY} now read ($q$ represents $u$ or $d$) 
\eqs{
\gamma_{Q}^\dagger&=&\gamma_{Q}\,, \quad  \gamma_{q}^\dagger = \gamma_{q}\,,\\
\sD Z_\Phi &=& 2 \gamma_\Phi Z_\Phi \,, \label{GPhi}\\
 \sD Z_{Q_L} &=& \gamma_{Q} Z_{Q_L}+Z_{Q_L} \gamma_{Q}\,,\label{AA}\\ 
 \sD Z_{q_R} &=& \gamma_{q}^* Z_{q_R}+Z_{q_R} \gamma_{q}^*\,,\label{BB}\\
 \ep\,Z_{Qq\Phi}+ \sD Z_{Qq\Phi} &=& Z_{Qq\Phi} \gamma_{q}^*+\gamma_{Q} Z_{Qq\Phi}+\gamma_\Phi Z_{Qq\Phi} \,. \label{CC}\quad 
}
The Hermiticity of anomalous dimensions has been used to simplify the remaining equations. 
Note that in the above equations $\sD$ acts only on dimensionless quantities, because  $\mu^{\ep}$ has been explicitly factored out from $Z_{Qq\Phi}$, 
and therefore one can  effectively set $\sD\simeq\beta^C\derp{}{g^C}$. 
I have solved these equations (together with their counterparts for SM leptons) up to the terms $\cO(\hbar^3)$. In the first step, 
I compared only the terms without the poles at 
$\epsilon=0$ to get 3-loop anomalous dimensions of fermions (and the Higgs field), as well as 3-loop beta functions for Yukawa matrices. To that end, I needed only the tree-level beta functions for other SM couplings. The so-obtained 3-loop beta functions are identical with the ones explicitly given in Ref. \cite{3L2HDMfiles} which relied on the adjustment of unitary factors. Moreover, the resulting anomalous dimensions of fermions are indeed Hermitian.  
Next, I have verified that these $\beta$'s and $\gamma$'s ensure that the pole parts on both sides of Eqs. \refer{GPhi}-\refer{CC} are the same, as expected. To check this, I needed also (respectively, 2-loop and 1-loop) $\beta$ functions for gauge and quartic couplings (which are given in Ref. \cite{3L2HDMfiles})
\footnote{In fact, those lower-order beta functions can be unambiguously obtained by demanding that the pole-terms on both sides of (counterparts of) Eqs.\refer{GPhi}-\refer{AA} for the Higgs doublet, gauge bosons and lepton doublets indeed agree, see the supplementary Mathematica notebook \cite{MNB}.}, as well as $\beta$ functions for gauge-fixing parameters. The latter, because of nonrenormalization theorems for gauge fixing terms in linear gauges \cite{ZJorig,ZJbook}, depend only on anomalous dimensions of the gauge bosons, and, analogously to \refer{GPhi}, can be easily obtained from the $Z_{B,W,G}$ factors listed in  \cite{3L2HDMfiles}. Finally, I compared the anomalous dimensions obtained by the proposed prescription, with the ones that follow from Eq. \refer{Eq:gammaNaive} \emph{after} the replacement $\sqrt{Z_F}\mapsto \mathcal{U}\sqrt{Z_F}$ with unitary factors $\mathcal{U}$ from Ref. \cite{3LSM}, and both approaches gave the same result. 

\vspace*{3pt}

%\section{Derivation of RGE}
\noindent {\bf Derivation of RGE.}
It remains to show that Eqs. \refer{DZF}-\refer{DZY} indeed guarantee that 
the renormalized 1PI generating functional $\Gamma$ obeys the RGE. To that end, 
I follow the approach originally used by Zinn-Justin \cite{ZJorig} to derive Slavnov-Taylor identities expressing the BRST symmetry of $\Gamma$ functional  in a gauge-symmetry-preserving regularization. Let $\{{g}^C\}$ be the set of independent renormalized parameters of a given theory, while  $\{\Psi^I\}$ -- the set of all (renormalized!) fields.  Suppose that the coefficients ${\beta}^C$ and $\gamma^I_{\ J}$ have been found which guarantee that the action $\cI$ that contains counterterms (i.e. generates renormalized correlation functions) obeys the equation 
\eq{ \label{RGEforI}
\mathscr{R}\cI=0,
} 
with the following differential operator 
\footnote{
The set $\{\Psi^I\}$ contains both, Weyl fermions $\chi$ and their conjugates 
$\bar\chi$. One can assume that anomalous dimensions of $\bar\chi$ are related by complex conjugation to those corresponding to $\chi$.
}   
\eq{\label{scrR}
\mathscr{R}\equiv\mu\derp{}{\mu} 
+
{\beta}^C\derp{}{g^C}
-\gamma^I_{\ J} \int{\rm d}^dx\, \Psi^J(x) \derf{}{\Psi^I(x)} \,.
}
For vertices explicitly shown in Eq. \refer{Eq:LagrWeylOg}, the identity  \refer{RGEforI} reduces to Eqs. \refer{DZF}, \refer{DZS} and \refer{DZY}. 
In particular, the existence of such ${\beta}^C$ and $\gamma^I_{\ J}$ functions follows from the arguments given below Eq. \refer{DZY}, as their generalization to other vertices is trivial once the well-known structure of counterterms of 
non-Abelian gauge theories \cite{ZJorig,ZJbook} in a gauge-symmetry-preserving regularization is taken into account.
\footnote{In fact, in \cite{Chank} an inductive proof was given that, in a variant of smooth cutoff regularization, the complete action (containing counterterms restoring finiteness as well as counterterms restoring BRST-invariance of correlation functions) also obeys Eq. \refer{RGEforI}. Thus, the arguments given 
here are not really restricted to gauge-symmetry-preserving regularizations.} 

Note that $\mathscr{R}$ is a first order partial differential operator and, importantly, the coefficients that multiply derivatives are at most linear in  quantum (i.e. propagating) fields. Thus, $\mathscr{R}$ belongs to the class of operators for which Zinn-Justin's trick \cite{ZJorig,ZJbook} works, i.e. by putting 
Eq. \refer{RGEforI} under the path integral and integrating by parts
one sees that the generating functional 
\eq{\nn
{\bf Z}[\cJ,g,\mu]=\int[\mathcal{D}\Psi]~\!
\exp\left\{i\,\cI+i\int{\rm d}^dx\,\cJ_I(x)\,\Psi^I(x)\right\}\, ,
}
obeys the following identity
\eq{\nn
R\,{\bf Z}[\cJ,g,\mu]
=\Theta \times {\bf Z}[\cJ,g,\mu]\,,
}
where
\eq{\nn
R\equiv \mu\derp{}{\mu} 
+
{\beta}^C\derp{}{g^C}
+\gamma^I_{\ J} \int{\rm d}^dx\, \cJ_I(x) \derf{}{\cJ_J(x)}\,,
}
and 
\eq{ \nn \phantom{aaaaa}
\Theta=\sum_I\, (\mp\gamma^I_{\ I})\times
\int{\rm d}^d x\, \delta^{(d)}_{\rm position}(0)\, ,
}
%
%\vspace*{3pt}
%
(upper/lower sign corresponds to bosonic/fermionic field $\Psi^I$). 
Since $\delta^{(d)}_{\rm position}(0)$ is a pure quartic divergence, 
it vanishes in DimReg; in any case, $\Theta$ disappears from the RGE 
satisfied by the  functional
$W$ that generates connected Green's functions provided it is defined by 
\cite{Chank}
\eq{\nn
\exp(iW[\cJ,g,\mu])
=\frac{{\bf Z}[\cJ,g,\mu]}{{\bf Z}[0,g,\mu]}\, ,
}
and thus 
\eq{\label{Eq:RGE-dla-W-bar}
R\,W[\cJ,g,\mu]=0\,.
}
It should be stressed that ${\bf Z}$ and $W$ generate renormalized correlation functions because the action $\cI$ already includes the counterterms. 
Performing the Legendre transform of $W[\cJ,g,\mu]$, it is now easy to check that the functional $\Gamma[\Psi,g,\mu]$ generating renormalized 1PI functions obeys 
the very same RGE as the action with counterterms, i.e. 
\eq{ %\label{RGEforI}
\mathscr{R}\Gamma[\Psi,g,\mu]=0,
} 
with $\mathscr{R}$ defined in \refer{scrR}. 
In particular, as I have already emphasized discussing finiteness of $\beta$ and $\gamma$ coefficients, the RGEs for `bare vertices' \refer{DZF}, \refer{DZS} and \refer{DZY} are structurally identical with the RGEs for the corresponding  
{renormalized} 1PI correlation functions.

\vspace*{3pt}
    
%\section{Conclusions}
\noindent {\bf Conclusions.}
The prescription given in the present paper allows for an efficient extraction of finite RGE coefficients \emph{directly} from countertems calculated in perturbation theory with diagrams involving only the standard set of (quantum) fields in a given model. It is based on solving a set of linear equations that can be most easily obtained by extracting independent vertices on the left hand side of Eq. \refer{RGEforI}, and requires no adjustment of extra unitary factors whatsoever. In fact, there is even no need to calculate square-roots of $Z$ matrices nor conventional `bare couplings', as everything is naturally expressed in terms of entities directly accessible in perturbation theory. 

The Hermiticity/symmetry conditions \refer{Herm}/\refer{Symm} ensure that the resulting system of linear equations has a unique solution (which, importantly, is finite). One can rephrase the arguments given above as follows. Since equations are linear, the unique solution  is guaranteed by non-vanishing determinant of the coefficient matrix. Since the determinant is a formal power series, it is non-vanishing because the system of equations has a unique solution \emph{at the tree-level}.

Nonetheless, conditions \refer{Herm}/\refer{Symm} are dictated by naturalness and simplicity, and without them the remaining equations have infinitely many solutions, reflecting the ambiguities in the approach based on adjusting unitary factors. In particular, I reproduced the original 3-loop beta functions of the SM  because the unitary factors chosen in Ref. \cite{3LSM} also lead to Hermitian anomalous dimensions of quarks.  

It is also worth noticing that, at the 3-loop order, flavor-improved RG coefficients used in Ref. \cite{Ambi} do not obey Hermiticity/symmetry conditions\footnote{It is so because the  coefficients calculated in Sec. 4.5 of Ref. \cite{Ambi} have non-vanishing non-pole terms.}, and therefore they represent a different solution to the problem considered in this paper, which is more suitable in certain situations (e.g. conformal field theories can be identified by vanishing of flavor-improved betas \cite{Fortin}). Still, simplicity of conditions \refer{Herm}/\refer{Symm} makes the RG coefficients obtained by imposing them a strong candidate for `the standard' beta and gamma functions.      

I~believe that the prescription given here can be useful for high-loop calculation of RGE coefficients in a generic renormalizable theory, and hope that it sheds some light on problems appearing in a more traditional approach.

\vspace{0.2cm}

\noindent{\bf Acknowledgments:} I am grateful to \mbox{Piotr~H.~Chankowski} for critical reading of an early version of the manuscript. I am also grateful to \mbox{Florian~Herren} and \mbox{Anders~Eller~Thomsen} for helpful correspondence regarding Refs. \cite{3L2HDM,3L2HDMfiles} and \cite{Ambi}.

%
%\onecolumngrid
%\vspace{0.9 cm}
%\begin{figure}[ht]
%\hspace{0.0cm}
%\begin{minipage}[b]{0.01\linewidth}
%\centering
%\includegraphics[scale=0.8]{d-aa}
%\label{fig:figure1}
%\caption{One-loop contributions to the vacuum polarization. 
%$\tig^{\mu\nu}_{\alpha\beta}(l)$.
%}
%\end{figure}
%\end{minipage}
%\hspace{7.7cm}

\end{document}